\documentclass[12pt, a4paper, parskip, DIV=15]{scrartcl}
\usepackage[english]{babel}
\usepackage[utf8]{inputenc}
\usepackage[T1]{fontenc}  
\usepackage{lmodern, textcomp}
\usepackage[hidelinks]{hyperref}
\usepackage[round]{natbib}
\usepackage{authblk, doi, graphicx}
\usepackage{grffile}  
\usepackage[separate-uncertainty=true]{siunitx}
\usepackage{amsmath, xspace}
\usepackage[format=plain, indention=0.5cm, font=footnotesize]{caption}
\usepackage{sidecap}
\usepackage{xcolor}

\catcode`\^^M=10

\newcommand{\ind}[1]{_\text{#1}}
\renewcommand{\vec}{\boldsymbol}
\newcommand{\parent}[1]{\ensuremath{\left(#1\right)}\xspace}
\newcommand{\fof}{\!\parent}

\newcommand{\md}{\ind{mod}}
\newcommand{\obs}{\ind{obs}}
\newcommand{\omM}{\omega\hspace{-0.1em}M}
\newcommand{\omMf}{\omM\fof f}
\newcommand{\M}[2]{\ensuremath{M\ind{#1#2}}\xspace}
\newcommand{\Ml}{\ensuremath{M\ind L}\xspace}
\newcommand{\Mw}{\ensuremath{M\ind w}\xspace}
\newcommand{\fc}{\ensuremath{f\ind c}\xspace}

\title{\LARGE Fast and robust earthquake source spectra and moment magnitudes from envelope inversion}
\author[1,*]{Tom Eulenfeld}
\author[2,3]{Torsten Dahm}
\author[3]{Sebastian Heimann}
\author[1]{Ulrich Wegler}

\affil[1]{Friedrich Schiller University Jena, Institute for Geosciences, Germany}
\affil[2]{GFZ German Research Centre for Geosciences}
\affil[3]{University of Potsdam, Institute for Geosciences}
\affil[*]{contact: tom.eulenfeld@uni-jena.de}
\date{{\vspace{-0.5cm}\small \today}}
\hypersetup{pdfauthor={Tom Eulenfeld et al.}, pdftitle={Robust earthquake spectra}}

\defcitealias{NetworkWebnet}{Institute of Geophysics, ASCR, 1991}

\begin{document}

\maketitle

\vspace{-1.5cm}
\begin{center}

{ \color{gray} \footnotesize
An edited version of this paper was published by\\\emph{Bulletin of the Seismological Society of America}, doi:\href{http://doi.org/10.1785/0120210200
}{10.1785/0120210200}.
}
\end{center}

\vspace{-0.5cm}
\begin{abstract}\noindent

With the present study we introduce a fast and robust method to calculate the source displacement spectra of small earthquakes on a local to regional scale.
The work is based on the publicly available \emph{Qopen} method of full envelope inversion which is further tuned for the given purpose.
Important source parameters -- seismic moment, moment magnitude, corner frequency and high-frequency fall-off -- are determined from the source spectra by fitting a simple earthquake source model.
The method is demonstrated by means of a data set comprising the 2018 West Bohemia earthquake swarm.
We report moment magnitudes, corner frequencies, and centroid moment tensors inverted from short period body waves with the \emph{Grond} package for all earthquakes with a local magnitude larger than 1.8. Moment magnitudes calculated by envelope inversion show a very good agreement to moment magnitudes resulting from the probabilisitc moment tensor inversion. Furthermore, source displacement spectra from envelope inversion show a good agreement with spectra obtained by multiple taper analysis of the direct onsets of body waves, but are not affected by the large scatter of the second.
The seismic moments obtained with the envelope inversion scale with corner frequencies according to $M_0 \propto f\ind c^{-4.7}$. Earthquakes of the present data set result in a smaller stress drop for smaller magnitudes. Self-similarity of earthquake rupture is not observed.
Additionally, we report frequency-dependent site amplification at the used stations.

\newline
\newline
Key points:
\begin{itemize}
\item We determine earthquake source spectra and derived parameters in a fast and robust manner
\item Comparison to moment tensor inversion and multiple taper spectrum analysis confirms robustness of the method
\item No self-similarity is observed for the 2018 West Bohemia earthquake swarm
\end{itemize}
\vspace{1ex}
\end{abstract}

\section{Introduction}
\label{sec:intro}

Local magnitudes are routinely reported, but usually other source parameters or source displacement spectra are not determined for small earthquakes (magnitude $\le$ 4) due to the additional effort involved. Earthquake source spectra have been determined earlier with envelopes calculated in different frequency bands \citep[e.g.][]{Mayeda1996} or by calculating spectra from direct wave arrivals \citep[e.g.][]{Hanks1972}. \cite{SensSchoenfelder2006a} suggested a physics-based model for estimating earthquake source spectra by inverting the envelope of recordings. The method was later refined by \cite{Eulenfeld2016}.\par

Several source parameters -- seismic moment, moment magnitude, corner frequency and high-frequency fall-off -- can be estimated from the source displacement spectrum.
Reporting source parameters routinely also for small earthquakes is beneficial to several fields in seismology. Moment magnitudes are based on the seismic moment which directly relates to physical source properties as the average slip and the area of the ruptured plane. They are important to derive adequate seismic hazard relationships. Corner frequencies of earthquake spectra are related to the dimension of the source and can be used -- together with seismic moment -- to calculate an estimate of the stress drop due to the rupture. The high frequency fall-off of source spectra is seldom analyzed, because on the one hand intrinsic attenuation and scattering has a larger impact for high frequencies and it is therefore difficult to estimate for small earthquakes and secondly, because the omega-square model \citep{Brune1970} with a high-frequency fall-off of~2 represents most earthquakes adequately well. But deviations from the omega-square model have been observed \citep[e.g.][]{Uchide2016, Eulenfeld2016}.
Therefore, the study of high-frequency spectral decay in earthquakes can help to unravel the controversy on different standard rupture models.
\par

In this article we propose a robust and fast scheme to invert full seismic envelopes for earthquake spectra with the Qopen method \citep{Eulenfeld2016}. The scheme allows us to simultaneously estimate frequency-dependent attenuation values and site amplification factors in addition to the earthquake source parameters. We apply the method to a mid-crustal earthquake swarm in the granitic Bohemian massive near Nový Kostel in 2018 \citep{Fischer2014}. Estimates of source parameters are compared to results of full waveform, probabilistic centroid moment tensor inversion from short period body waves using the Grond software \citep{Grond}. The resulting source spectra are additionally compared to the source spectra estimated from direct onsets of body waves. After the validation of the method, we present source parameters and conclusions related to the 2018 West Bohemia earthquake swarm and discuss further implications of the method for quantifying site effects and attenuation values.

\section{Data and method}

\subsection{2018 West Bohemia earthquake swarm}

\label{sec:data}

\begin{SCfigure}
\centering
\includegraphics[width=0.5\textwidth]{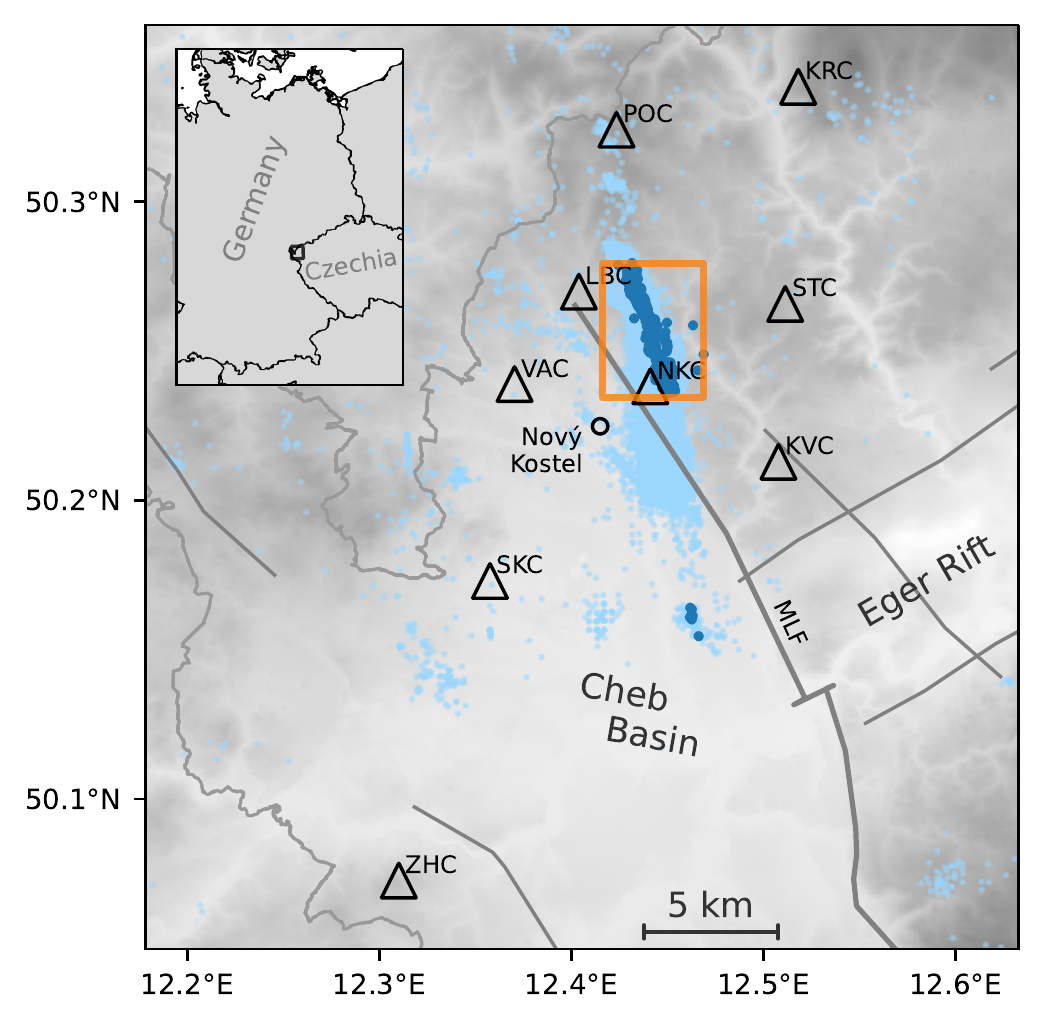}
\caption{Topographic map of Czech-German border region near Nový Kostel. Displayed are WEBNET seismic stations used in this study (triangles) together with local seismicity between the years 1994 and 2021 according to WEBNET catalog (light blue) and epicenters of 2018 Nový Kostel swarm earthquakes (dark blue). The orange rectangle defines the scope of the map in figure~\ref{fig:eventmap}. The Mariánské Lázně fault zone (MLF) and other tectonic lineaments are indicated with gray lines.
}
\label{fig:map}
\end{SCfigure}

\begin{figure}
\centering
\includegraphics[width=0.9\textwidth]{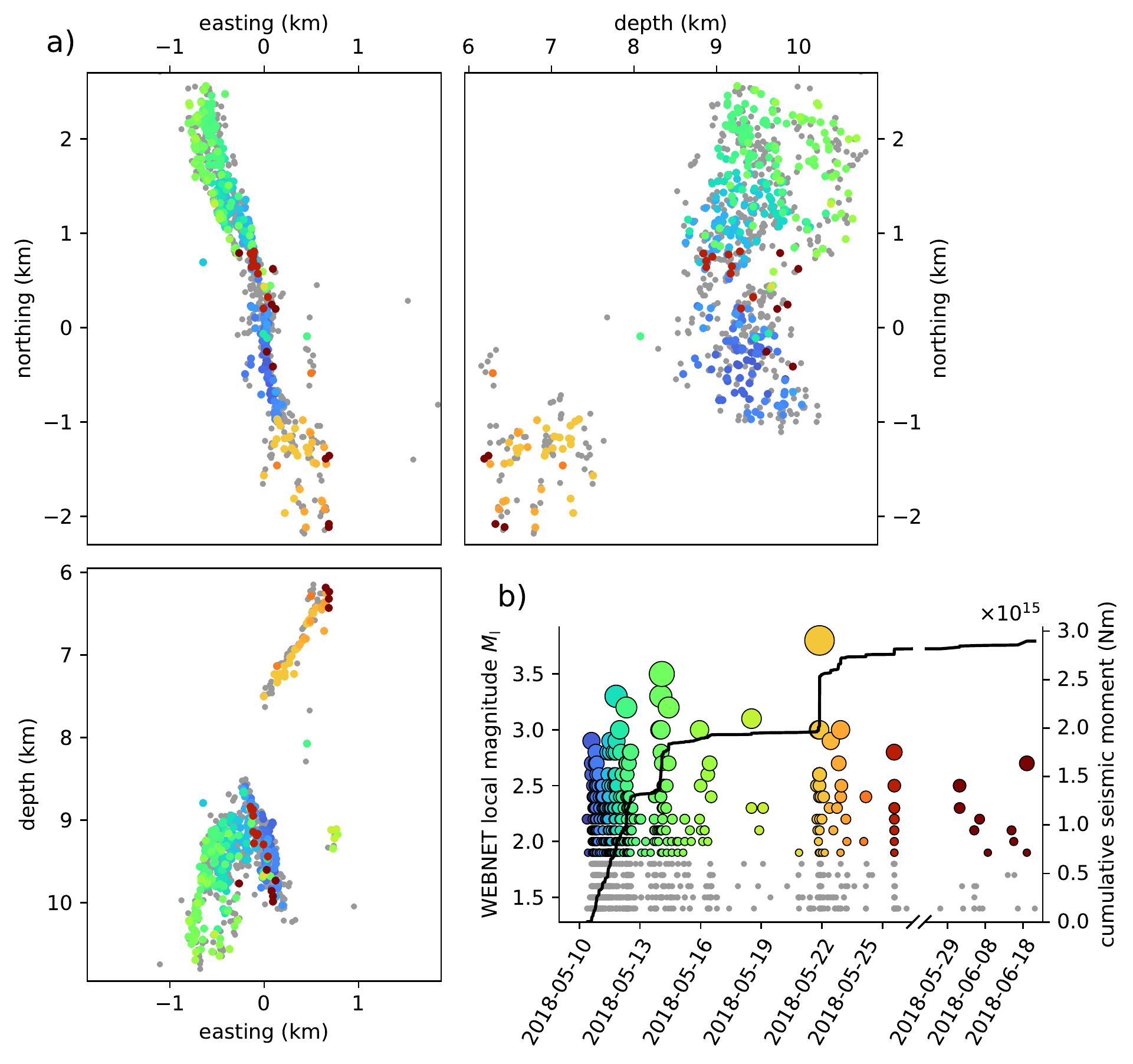}
\caption{a) Map and depth sections of all 966 events of the 2018 earthquake swarm \citep[gray,][]{Bachura2021}. 376 events have a magnitude larger than 1.8 and are color-coded with time.
Coordinates in the map are relative to 50.25°N, 12.45°E.\newline
b) Local magnitude and cumulative seismic moment of the same earthquakes plotted versus time.
The underlying seismic moments are determined in this study -- either directly from envelope inversion or indirectly from local magnitude with the obtained magnitude relationship.
}
\label{fig:eventmap}
\end{figure}

The topography of the Czech-German border region West Bohemia / Vogtland is shown in figure~\ref{fig:map} together with used WEBNET seismic stations and earthquake epicenters between the years 1994 and 2021.
The double difference catalog of the 2018 swarm used in this study was compiled by \cite{Bachura2021} and is displayed in figure~\ref{fig:eventmap}a. It consists of approximately 1000 earthquakes with local magnitudes larger than 1.3. The uncertainties in origin locations relative to each other are \SI{50}m. For the purpose of this study, we use the events of a curated catalog for which coda does not interfere with other earthquakes. This catalog is available from \cite{Eulenfeld2020_sourcecode, Eulenfeld2020} and embraces around 150 earthquakes with local magnitudes larger than 1.8.
The activity started at May~10, 2018 at a depth of \SI 9{km} to \SI{10}{km} and migrated to the north after May~11.
On May~21 earthquakes started rupturing a region south of the previous activity at shallower depth (around~\SI 7{km}). After June~19 the activity faded out.
Figure~\ref{fig:eventmap}b displays the distribution of local magnitudes versus time together with the cumulative seismic moment.
In figure~\ref{fig:eventmap}b the activity shows a gradual transition from swarm stage to mainshock/aftershock stage (May~21) which was originally reported by \citet{Bachura2021}.
The waveforms of the earthquakes were registered on the Czech WEBNET stations with \SI{250}{Hz} sampling rate. For this study, data from 9 WEBNET stations are used (figure~\ref{fig:map}). The earthquake catalog, station metadata and waveform data are available at \url{https://doi.org/10.5281/zenodo.3741464} \citep{DatasetEQSwarm2018}.

\subsection{Earthquake source spectra from full envelope inversion}
\label{sec:method1}

We use the Qopen method \citep{Eulenfeld2016} to separate intrinsic and scattering attenuation by envelope inversion and to estimate the source displacement spectra. The method of full envelope inversion was introduced by \cite{SensSchoenfelder2006a} and was further developed and implemented by \cite{Eulenfeld2016, Eulenfeld2017}.
The method is comprehensively described in \cite{Eulenfeld2016} and therefore only the most important points are recapitulated here. Enhancements related to the real-time determination of earthquake source spectra are introduced by this study. We use the dedicated scripts at \url{https://github.com/trichter/qopen} \citep{Qopen} and incorporated our improvements.\par

The observed energy density envelopes $E\obs$ are calculated from the 3 component restituted, filtered seismic velocity records $\dot u_c$ with the help of the Hilbert transform $\mathcal H$ (\citealp{Sato2012}, page~41; \citealp{Eulenfeld2016}, equations 3--4)

\begin{equation}
E\obs\fof{t, \vec r} = \frac{\rho\sum_{c=1}^3\parent{\dot u_c\fof{t, \vec r}^2 + \mathcal H\fof{\dot u_c\fof{t, \vec r}}^2}}{2C\ind{energy}\Delta f} 
\label{eq:Eobs}
\end{equation}

with the mean mass density $\rho$, energy free surface correction $C\ind{energy}{=}4$ \citep{Emoto2010} and filter bandwidth $\Delta f$ \citep[e.g.][]{Wegler2006a}. The central frequencies of the bandpass filter are chosen between \SI{0.75}{Hz} and \SI{96}{Hz}. These observed energy densities are compared to synthetic envelopes which are given by equation (1) in \cite{Eulenfeld2016}

\begin{equation}
E\md\fof{t, \vec r} = WR\fof{\vec r}G\fof{t, \vec r, g}e^{-bt} \,. \label{eq:Emod}
\end{equation}

$W$ is the spectral source energy of the earthquake, $R\fof{\vec r}$ is the energy site amplification factor at the stations. $e^{-bt}$ describes the exponential intrinsic damping with time and depends on the intrinsic absorption parameter $b$. The Green's function $G\fof{t, \vec r, g}$ with scattering strength $g$ accounts for the direct wave and the scattered wave field and is given by the approximation of the solution for 3-dimensional isotropic radiative transfer of \cite{Paasschens1997}.

Because the inversion is performed in different frequency bands, $W$, $R$, $g$ and $b$ will be determined as a function of frequency $f$. The spectral source energy $W\fof f$ is converted to the S~wave source displacement spectrum $\omMf$ with equation~11 of \cite{Eulenfeld2016} \citep[page~188]{Sato2012}

\begin{equation}
  \omMf = \sqrt{\frac{5 \rho v\ind S^5 W\fof f}{2\pi f^2}} \label{eq:sds}
\end{equation}

with mean S~wave velocity $ v\ind S$. For mean density and mean velocity we use values of $\rho=\SI{2600}{kg/m^3}$ and $v\ind S=\SI{3.4}{km/s}$. These values together with $v\ind P=\SI{5.78}{km/s}$ for P~waves will also be used in section~\ref{sec:method3}.

The source displacement spectrum can be fitted by a source model of the form 

\begin{equation}
  \omMf = M_0 \parent{1+\parent{\frac f{f\ind c}}^{\gamma n}}^{-\frac 1\gamma} \label{eq:sourcemodel}
\end{equation}

with seismic moment $M_0$ and corner frequency $f\ind c$ \citep{Abercrombie1995}. $n$ is the high frequency fall-off and the shape parameter $\gamma$ describes the sharpness of the transition between the constant level $M_0$ for low frequencies and the fall-off with $f^{-n}$ for high frequencies. $n{=}2$ in equation~\ref{eq:sourcemodel} corresponds to the source displacement spectrum of an omega square model.
The equation system

\begin{equation}
\ln E\obs\fof{t, \vec r} = \ln E\md\fof{t, \vec r}  \label{eq:system}
\end{equation}
is solved in a least-square manner.

In this study we use a direct S wave window (\SI{-0.5}s, \SI{2.5}s) relative to the S onset and a coda window starting at the end of the direct wave window and extending until \SI{50}s after S onset. The coda window is shorter if the signal-to-noise ratio (SNR) falls below 2 or if the envelope increases with time due to nuisances. The envelope in the direct S wave window is averaged to mitigate the effect of forward scattering \citep{Eulenfeld2016}. This average is weighted according to the length of the direct S wave window.\par

The inversion for earthquake source spectra is performed in three steps:

\begin{description}
\item{1. Estimate intrinsic and scattering attenuation}\newline
Equation system (\ref{eq:system}) is solved for $W$, $R$, $g$ and $b$ for all frequency bands and all earthquakes separately. Because (\ref{eq:system}) is not linear in $g$, this inversion is itself an iterative process described in \cite{Eulenfeld2016}. $g$ and $b$ are averaged in each frequency band for different events in a robust manner and can be converted to Q values by equation~(15) in \cite{Eulenfeld2016}. Because $g$ and $b$ are medium properties, both values can be fixed for the following two steps. For this step we use only 39 earthquakes with a magnitude larger than 2.5 to guarantee a coda which is long enough to separate the two attenuation mechanisms.
\item{2. Refine and align station site amplification}\newline
Equation system (\ref{eq:system}) is again solved for $W$ and $R$ using the fixed values $g$ and $b$ determined in step~1 for all frequency bands and all earthquakes separately. Because of the co-linearity of $R$ and $W$ in equation~(\ref{eq:Emod}) and because each earthquake might be registered at a different set of stations, the site amplifications $R$ are re-aligned as described in section~2.2 of \cite{Eulenfeld2017}. Therefore the site amplification of a single station (e.g.\ a station with known or low site amplification) or the geometric mean over site amplifications of all stations needs to be fixed. In this study the geometric mean is fixed at 1. The $R$ values are geometrically averaged in each frequency band for each station for the final step.
\item{3. Calculate source displacement spectra, source parameters, moment magnitudes}\newline
Equation system (\ref{eq:system}) is again solved for a single value $W$ for each frequency band and earthquake using the fixed values $g$, $b$ and $R$ determined in steps 1 and 2. This step can use the same data set as in the previous steps, but also new earthquakes in the same region can be processed in a robust and fast manner without re-determination of the attenuation parameters and the site responses. Finally, spectral source densities are converted to source displacement spectra with equation~(\ref{eq:sds}) and source parameters can be determined by fitting the source model in equation~(\ref{eq:sourcemodel}).
\end{description}

\begin{figure}
\centering
\includegraphics[width=0.8\textwidth]{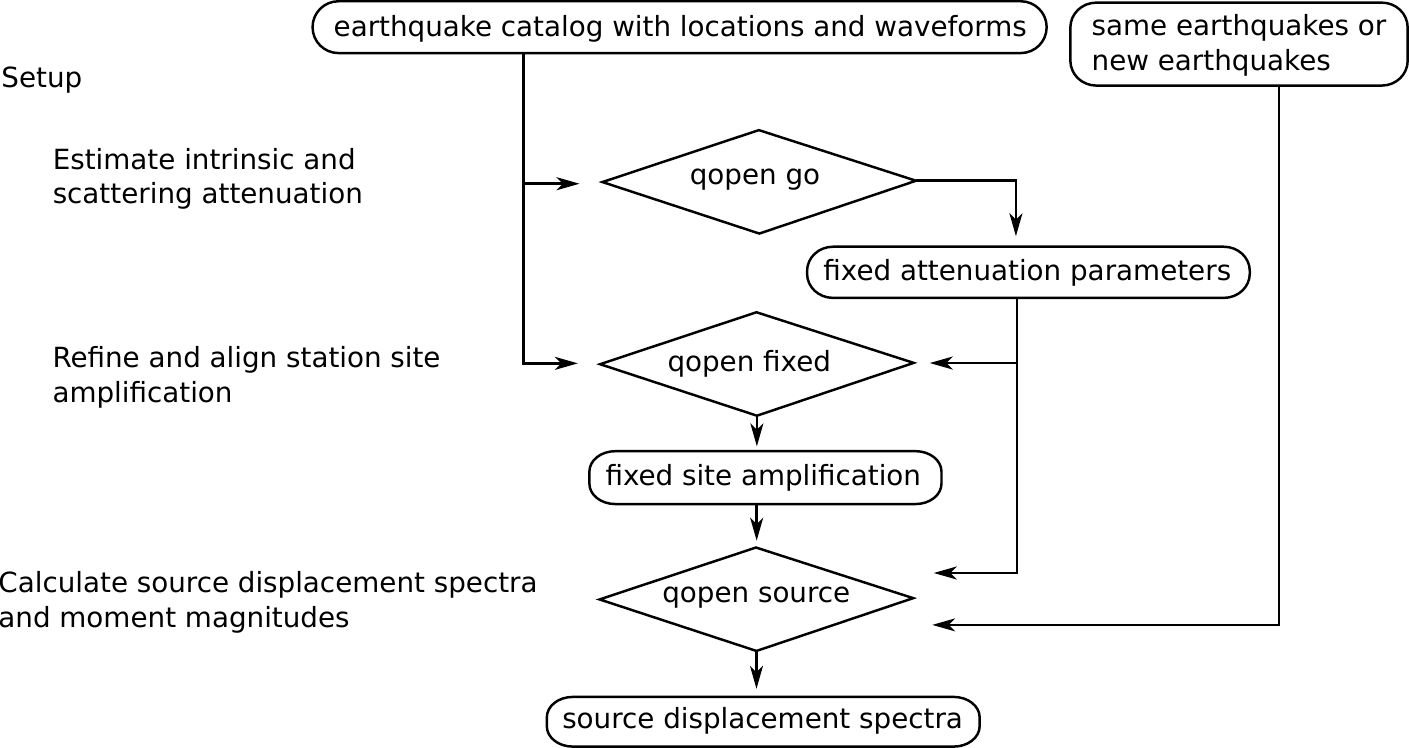}
\caption{Flow chart illustrating the invocation of Qopen commands to calculate source displacement spectra in a fast and robust manner.}
\label{fig:flow}
\end{figure}

Step~3 and the first part of step~2 are introduced by this study. Source displacement spectra can be equally derived in step 1 \citep{Eulenfeld2016, Eulenfeld2017, Eken2019}, but the adapted procedure guarantees a robust and fast determination of source parameters of previously not analyzed earthquakes after the initial setup in steps 1 and 2. Each of the above steps can be easily applied by a dedicated command in the Qopen scripts (figure~\ref{fig:flow}). Example fits for a single earthquake and frequency band are displayed in figure~\ref{fig:qopen_fits}.

\subsection{Moment tensors from waveform inversion}
\label{sec:method2}
Results of the method described in the previous section are compared to moment tensors estimated using a Bayesian bootstrap optimization
(BABO, implemented in the Grond software package, \citealp{Pyrocko,Grond})

The moment tensor inversion minimizes the residuals ($L_1$-Norm) between three-component displacement seismograms and synthetic Green's functions in a restricted space of centroid locations of earthquake point source models.
Forward modeling of synthetic seismograms is performed using a precomputed Green's function store \citep[GFDB,][]{Fomosto} calculated using an orthonormal propagator method for layered earth structures \citep[QSEIS program,][by Pyrocko implementation \emph{fomosto}]{Wang1999}. We used the GFDB \emph{vogtland\_scatter\_v4} available on the Pyrocko Green's Mill platform \url{https://greens-mill.pyrocko.org}. The velocity structure \citep[modified model from][]{Malek2005} is relatively smooth, which has been shown to be beneficial for body wave inversion of local earthquakes. The quality factor for S~waves is set to $Q\ind S{=}88$ corresponding to the total attenuation at a frequency of \SI{2.1}{Hz} obtained in section~\ref{sec:method1}. Total attenuation is given by the sum of contributions of intrinsic attenuation and scattering: $Q\ind S^{-1} = Q\ind{intr}^{-1} + Q\ind{sc}^{-1}$. While the ratio of quality factors of P and S~waves is often assumed to be $Q\ind P{/}Q\ind S{=}2.25$ for waves of low frequency, \citet[chapter~5.1]{Sato2012} point out that for frequencies higher than \SI 1{Hz} $Q\ind P{/}Q\ind S$ is usually lower than 1. For the sake of simplicity, we assume $Q\ind P{=}Q\ind S$ in the upper crust.

Along with the waveform, first-arrival times are calculated (Pyrocko ray tracer {\em cake}) and stored in the GFDB. These are later used to cut out appropriate P and S~wavelets per iteration in BABO. 

The observed three-component waveforms were corrected for instrument response and sensor orientation and converted to displacements. We applied the same acausal filters and tapers to observed and synthetic waveforms. Based on the SNR in the seismograms and the given magnitude range, a filter between 1 and \SI 3{Hz} was used (flat part of the filter response, dropping to zero at \SI{0.7}{Hz} and \SI{4.5}{Hz}). Waveforms were tapered to \SI{\pm 0.25}s around the onsets of the P and S~waves (fade-out over \SI{0.25}s), with the P~waves inverted on the vertical (Z) and the S~waves inverted only on the transverse components (T). Phase shifts of the onset times were allowed up to \SI{0.15}s, with a penalty function becoming active for nonzero phase shifts. The inversion scheme is similar to its application to small magnitude crustal earthquakes near Halle and Leipzig, several hundred kilometers north of the 2018 swarm \citep{Dahm2018}.

Optimization in BABO explores the entire model space and maps model parameter trade-offs with a flexible design of objective functions.  BABO explores a set of perturbed objective functions simultaneously for the model space regions that cover minima. Ensembles of well-fitting models are obtained, representing a non-parametric estimate of the posterior parameter probabilities along with the best solutions. In our application, the centroid origin time, location, and depth are searched, along with five independent moment tensor components. Automatic weights are applied to all observations to balance their contributions to the overall misfit. The weights are determined prior to optimization from forward modeling of an ensemble of 5000 random sources and computing misfits of synthetic traces against zero-amplitude observations \citep{Heimann2011_phd}. The reciprocals of the mean values of these misfit contributions are used as weights.

\par
\begin{figure}
\centering
\includegraphics[width=0.9\textwidth]{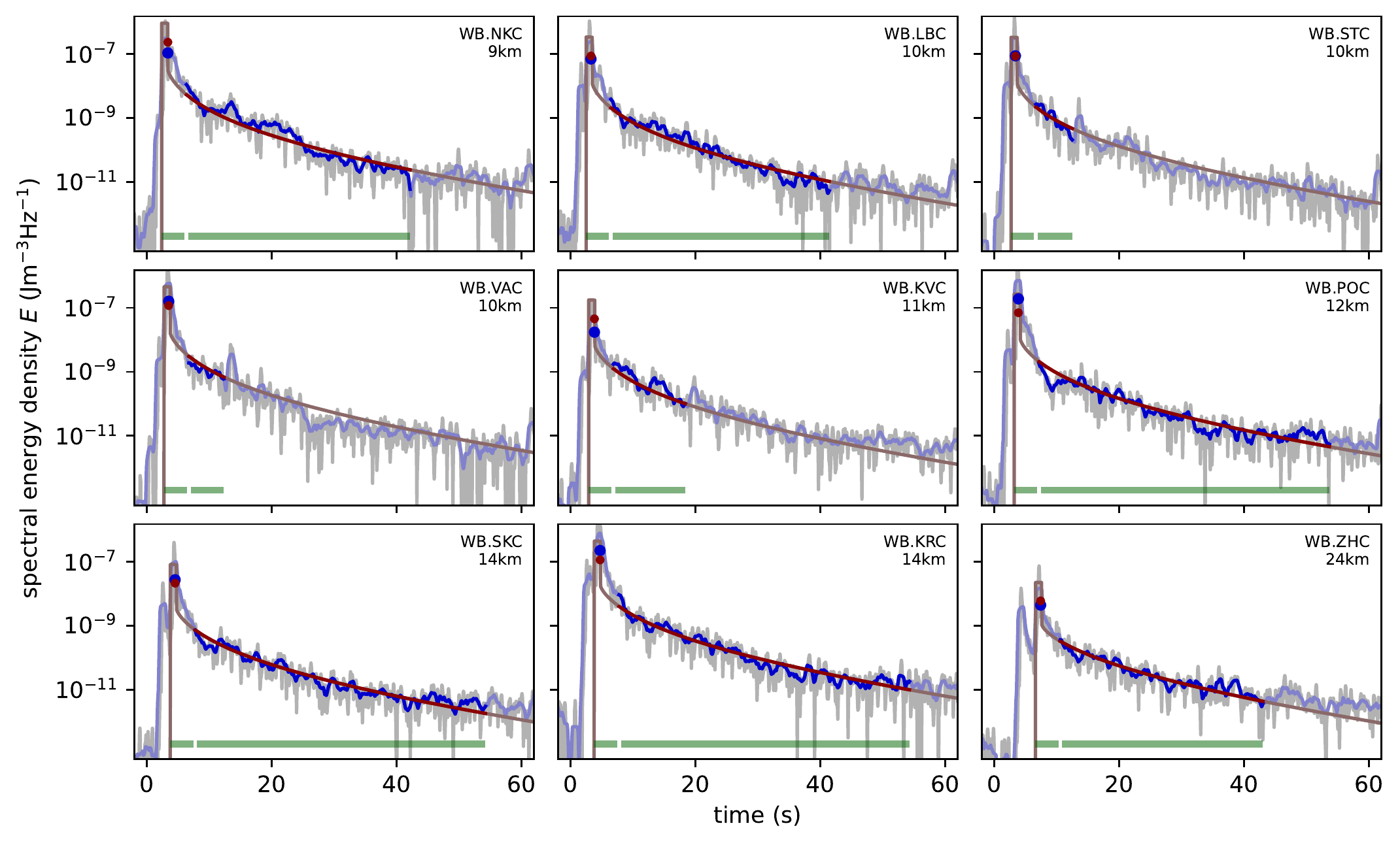}
\caption{Observed and synthetic envelopes (spectral energy density) for each station in a separate panel in the frequency range (\SI 4{Hz}, \SI 8{Hz}) calculated with Qopen for event 20186784.
Observed envelopes are displayed with gray lines, smoothed observed envelopes with blue lines and synthetic envelopes with red lines. The green bars in the bottom mark the two time windows used in the inversion. Envelopes in the earlier direct S wave time window are averaged (blue and red dots), envelopes in the coda time window are displayed as dark blue and dark red lines.
}
\label{fig:qopen_fits}
\vspace{0.5cm}
\includegraphics[width=0.9\textwidth]{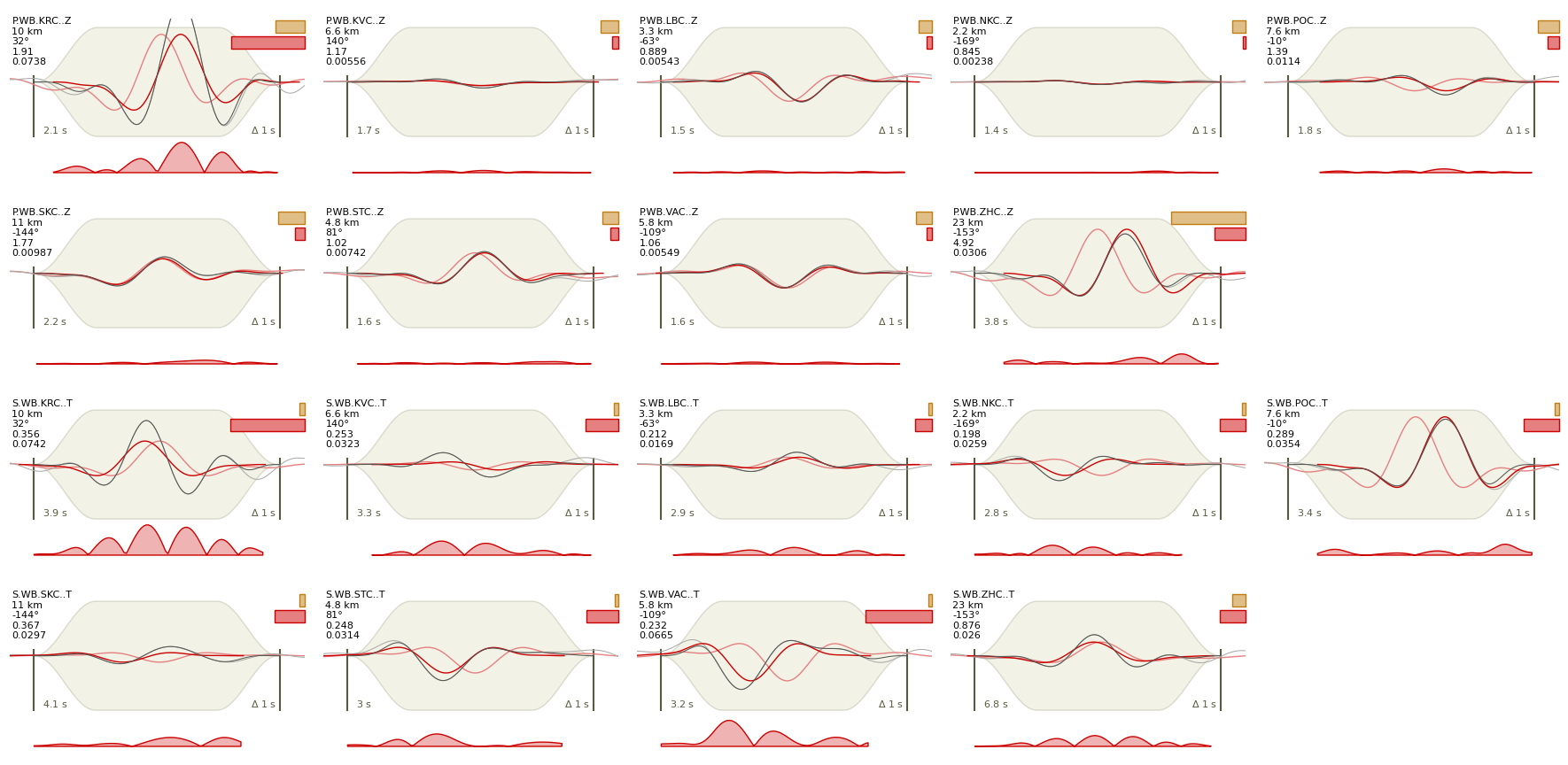}
\caption{Observed and synthetic waveforms calculated with Grond for the same event 20186784 as in figure~\ref{fig:qopen_fits}. The Z component is fitted in a \SI 1s long time window around P-onset (top 9 panels) and the T component is fitted in a \SI 1s long time window around S onset for each station. The observed waveforms are displayed in black, modeled waveforms in light red, modeled and shifted waveforms in red. The residuals are displayed in the bottom of each panel. The bars in each panel indicate the weight (yellow) and the misfit (red) of the corresponding fit. Please note the completely different features and time windows used in the Qopen and Grond inversions.}
\label{fig:grond_fits}
\end{figure}

From the moment tensor optimization, we obtain 148 solutions of which most are very stable 
(interactive reports under \url{https://data.pyrocko.org/publications/grond-reports/west-bohemia-2018/}, \citealp{Eulenfeld2021_grondreport}).
A waveform fit example is shown in figure~ \ref{fig:grond_fits}.
Qualitatively and judged by experience, the waveforms fit reasonably well for the whole range of magnitudes analyzed. Quantitatively, parameter error estimates obtained from the bootstrap confirm that they only weakly depend on magnitude. Median formal errors obtained through the bootstrap are: magnitude: 0.04, time: \SI{0.08}s, depth: \SI{390}m, north: \SI{450}m, east: \SI{560}m. Focal mechanism orientations of the solution ensemble scatter with a median Kagan angle \citep{Kagan1991} of \SI 7{\degree} around the mean solutions.
Trade-offs between source components are almost absent, with an exception between depth and time as a later origin time can be explained by a slightly shallower depth.

\subsection{Earthquake source spectra from direct body waves}
\label{sec:method3}
Grond does not invert for the full source displacement spectrum, but rather for seismic moment and focal mechanism.
Therefore, we additionally calculate source displacement spectra from the direct P and S~wave arrivals to compare these with source displacement spectra calculated with Qopen.
The processing starts with a restitution to displacement. Time windows (\SI{-0.1}s, \SI{0.5}s) around P- and S-pick are used to calculate the displacement spectrum with the mtspec library using multitaper spectrum analysis \citep{Thomson1982, Prieto2009, mtspec}. The spectrum is calculated for the vertical Z~component for the P~wave window respective the tangential T~component for the S~wave window. The displacement spectrum of P~waves $U\ind P$ is converted to source displacement spectrum $\omM\ind P$ with the formula

\begin{equation}
\omM\ind P\fof f = \frac{4\pi\rho r v\ind P^3 U\ind P\fof f}{C\ind{ampl} \mathcal R\ind P\fof{\vec r}} e^{\pi t f Q\ind P^{-1}} \,. \label{eq:sds_from_onset}
\end{equation}

$r$ is the distance between earthquake and station, $v\ind P$ the P~wave velocity, $C\ind{ampl}{=}2$ the amplitude surface correction. The radiation pattern $\mathcal R\ind P\fof{\vec r}$ is calculated from the corresponding focal mechanism determined in section~\ref{sec:method2}.
The term $e^{\pi t f Q\ind P^{-1}}$ compensates the damping by intrinsic and scattering attenuation.
Formula~\ref{eq:sds_from_onset} is given in \cite{Hanks1972} who refer to \cite{KeilisBorok1960}. Here only free surface correction and damping was added.

The source displacement spectrum from the S wave $\omM\ind S\fof f$ is calculated correspondingly. 
We again assume $Q\ind P{=}Q\ind S$ and calculate the frequency-dependent total attenuation from the contributions of intrinsic attenuation and scattering obtained in section~\ref{sec:method1}.
The correction for attenuation lead to an approximately \SI{25}{\%} higher source displacement spectrum at low frequencies depending on distance between the locations of source and station. 

Because we chose a window length of \SI{0.6}s, the frequency sampling in the Fourier domain and the smallest resolvable frequency is \SI{1.67}{Hz}.
For determining the low-frequency plateaus of the spectra we averaged the spectral amplitudes for frequencies smaller than \SI{3.5}{Hz}, corresponding to two data points in the frequency domain.
The median of those observations for different stations and wave types results in the seismic moment from onsets for each earthquake.

\begin{figure}
\centering
\includegraphics[width=0.98\textwidth]{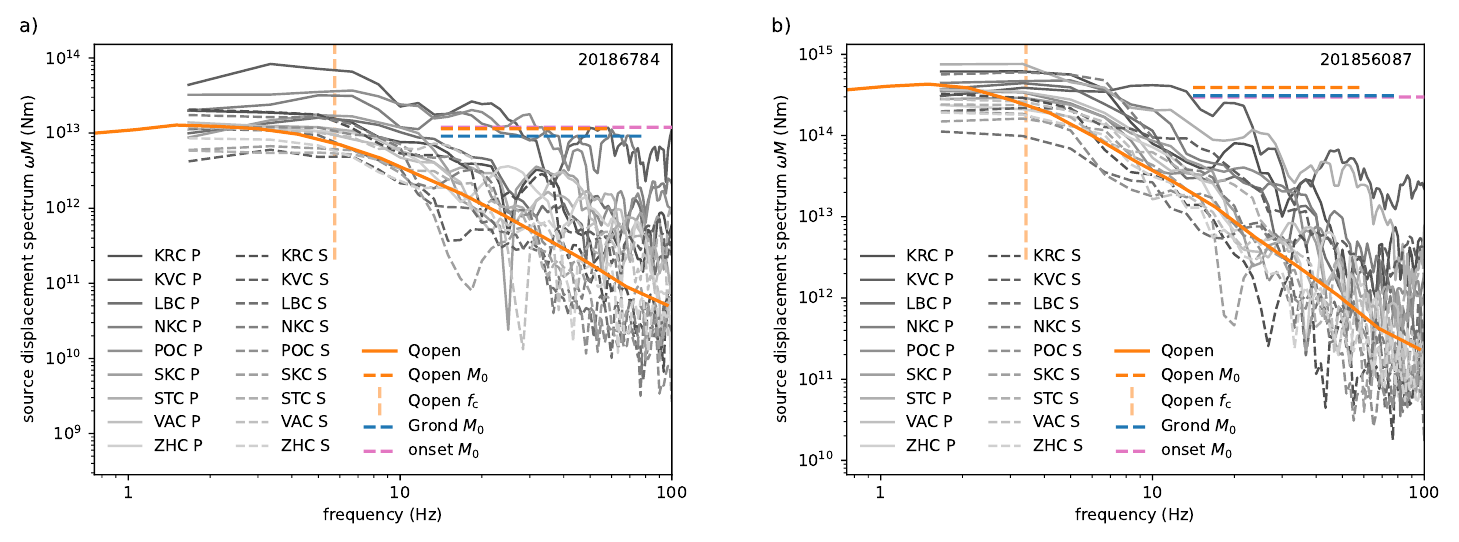}
\caption{Source displacement spectra calculated from the direct onset of P and S waves at the different stations (gray dashed and continuous lines) for two different events, 20186784 as in figure~\ref{fig:qopen_fits} ($\Ml{=}2.8$, left panel a) and the largest event 201856087 ($\Ml{=}3.8$, right panel b). The spectra are corrected for the radiation pattern and attenuation of the direct wave. The corresponding seismic moments calculated from the direct onsets are displayed as purple horizontal lines, seismic moments from Grond are displayed with blue lines and the source displacement spectra from Qopen and derived source parameters with orange lines. A good agreement in seismic moment is observed for the three different methods.}
\label{fig:sds_comp}
\end{figure}

Exemplary, spectra calculated with this method are displayed for two events in figure~\ref{fig:sds_comp}, together with the determined seismic moment. 
Results from the other two methods -- source displacement spectrum, seismic moment, corner frequency for Qopen and seismic moment for Grond -- are also displayed for comparison. The source displacement spectra for the different onsets show a large scatter in amplitude between different stations and as a function of frequency even after the dedicated correction for the radiation pattern. Contrary, the spectrum estimated from the envelope inversion is a relatively smooth function of frequency, presumably because coda waves present in the envelope have a more stable spatial pattern.
Still, source displacement spectra determined with Qopen and the source displacement spectra calculated from the wave onsets show a reasonable agreement.
A good conformity between the seismic moments of the different methods can also be noticed, this matter will be further analyzed in section~\ref{sec:mags}.

\section{Results}

\subsection{Comparison of scattering strength and intrinsic attenuation to previous studies}
\label{sec:Q}

The results of shear wave scattering strength and intrinsic attenuation are not the main goal of this study. They are necessary prerequisites to successfully fit the envelope observations and obtain source displacement spectra without any prior knowledge. Therefore a comparison to previous studies is reasonable.

\begin{figure}
\centering
\includegraphics[width=0.8\textwidth]{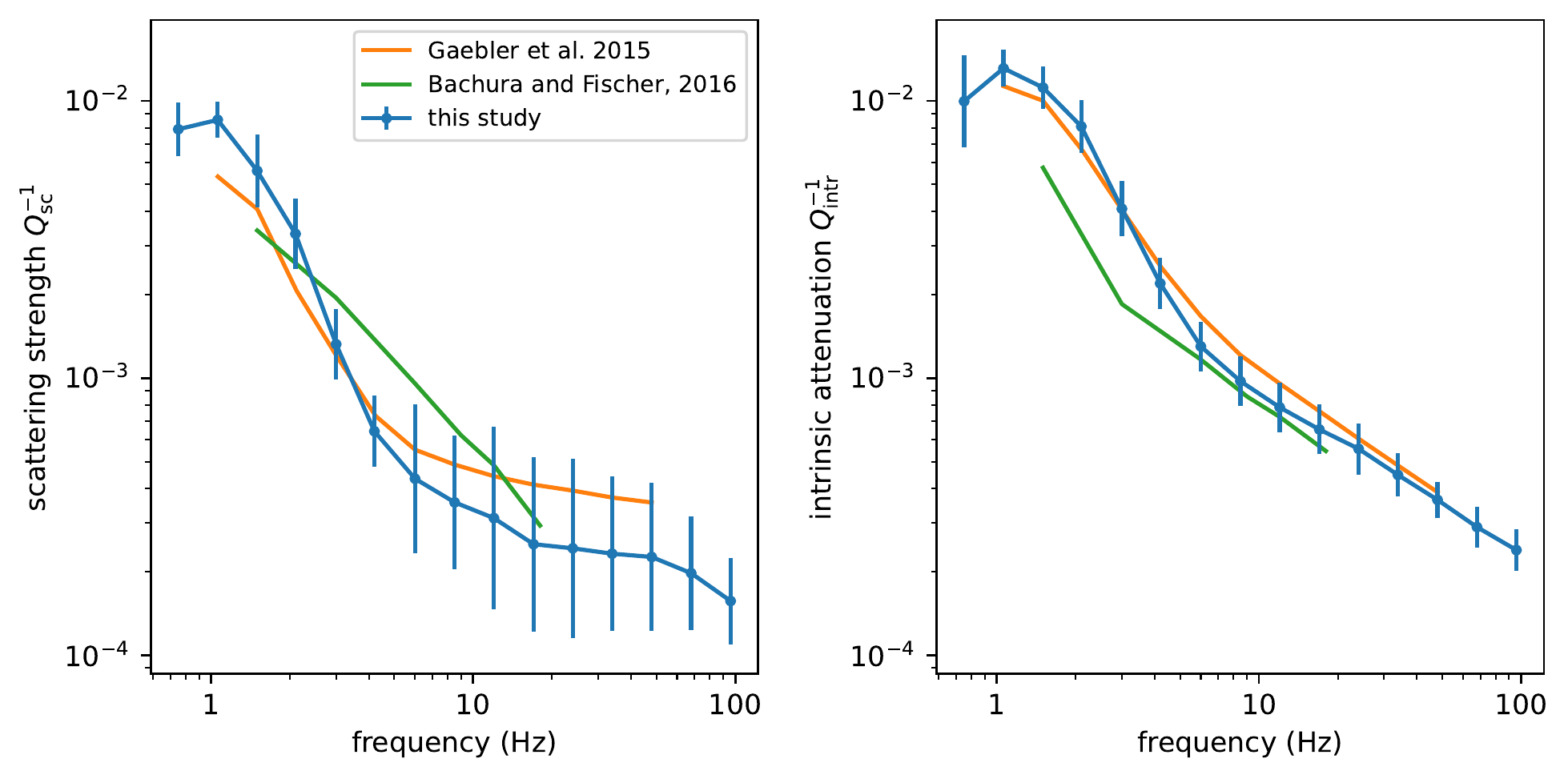}
\caption{Scattering strength and intrinsic attenuation as a function of frequency obtained in this study and compared to findings of \citet{Gaebler2015} and \citet{Bachura2016}.}
\label{fig:Q}
\end{figure}

\cite{Gaebler2015} performed an analysis with the same method for a larger region covering the study area. \cite{Bachura2016} used Multiple Lapse Time Window Analysis \citep[MLTWA,][]{Fehler1992, Hoshiba1993} to obtain the attenuation parameters in a similar setting using the 2011 earthquake swarm in West Bohemia. In figure~\ref{fig:Q} results of these studies are compared to our findings. Both previous studies did not include the whole frequency band used in our study. Anyway, a good agreement in amplitude and dependency on frequency of attenuation parameters is obvious between the different studies (figure~\ref{fig:Q}). $Q^{-1}$ values both for scattering and intrinsic attenuation range from $10^{-2}$ at \SI 1{Hz} to $10^{-4}$ at \SI{100}{Hz} with intrinsic attenuation dominating over scattering attenuation.
Results of both Qopen and MLTWA inversions performed by \cite{Laaten2021} indicate a weaker intrinsic attenuation below $Q\ind{intr}^{-1}=\num{e-3}$ and a weaker scattering strength below $Q\ind{sc}^{-1}=\num{3e-4}$ in a region extending from the present study area up to hundred kilometer to the north.
Several other studies determined values for total attenuation $Q^{-1} = Q^{-1}\ind{sc} + Q^{-1}\ind{intr}$.
\cite{Kriegerowski2019} obtained total shear wave attenuation values of \num{2e-3} to \num{4e-3} in the source region of the 2018 earthquake swarm in West Bohemia for frequencies above ${\approx}\SI{10}{Hz}$. Their results showed a high variation even with negative Q values for individual measurements, but have the same order of magnitude as our results. 
\cite{Haendel2019} used deconvolution of ambient noise recordings of a borehole and a co-located surface station to derive the near surface total shear wave attenuation in a frequency band between \SI 5{Hz} to \SI{15}{Hz}. Their $Q^{-1}\ind S$ values range between \num{e-2} and \num{e-1} and reflect the higher attenuation near the surface.

\subsection{Source displacement spectra and focal mechanisms}
\label{sec:params}

\begin{figure}
\includegraphics[width=0.9\textwidth]{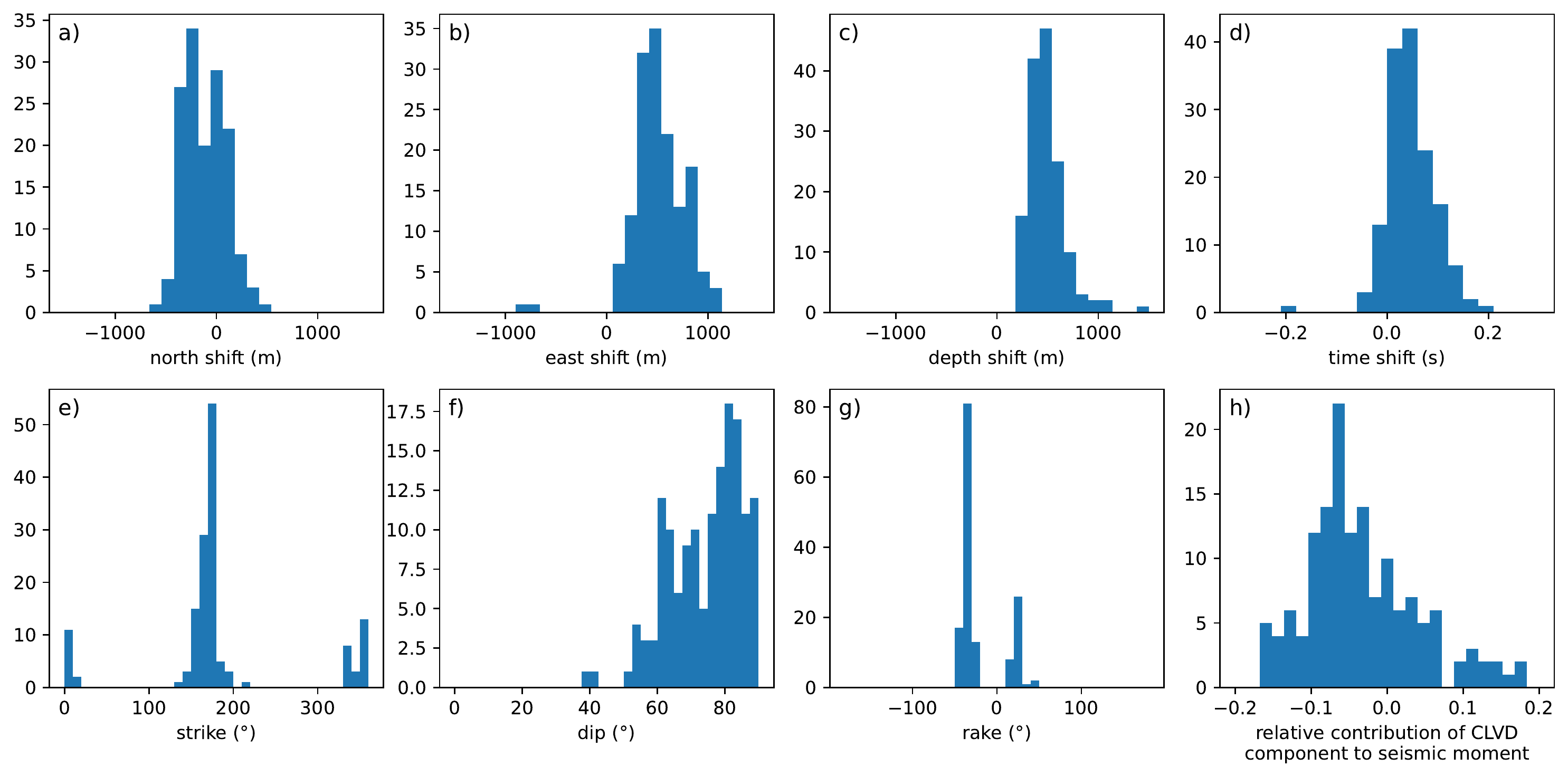}
\caption{a-d) Deviation of Grond's mean solution to catalog parameters. Events are located slightly deeper and more easterly relative to catalog locations, median deviation of origin time is less than \SI{0.1}s.\newline
e-g) Distribution of strike, dip and rake of the fault plane solution determined with Grond. Note, that strike and rake distributions appear bi-modal due to the steep dip.\newline
h) Distribution of contributions of compensated linear vector dipole (CLVD) component to seismic moment determined with Grond.
}
\label{fig:grond_residuals}
\end{figure}

Both Qopen and Grond invert the data for seismic moment, but the other targets are different. While Qopen additionally inverts for the source displacement spectrum, Grond inverts for the centroid location and focal mechanism. The centroid locations are systematically shifted by a median of about \SI{500}m to the east and larger depth, but deviations from catalog locations are in general below \SI 1{km}.
Deviations of origin time from catalog origin time are below \SI{0.2}s, its median is below \SI{0.1}s (figure~\ref{fig:grond_residuals}).
\par

\begin{figure}
\centering
\includegraphics[width=\textwidth]{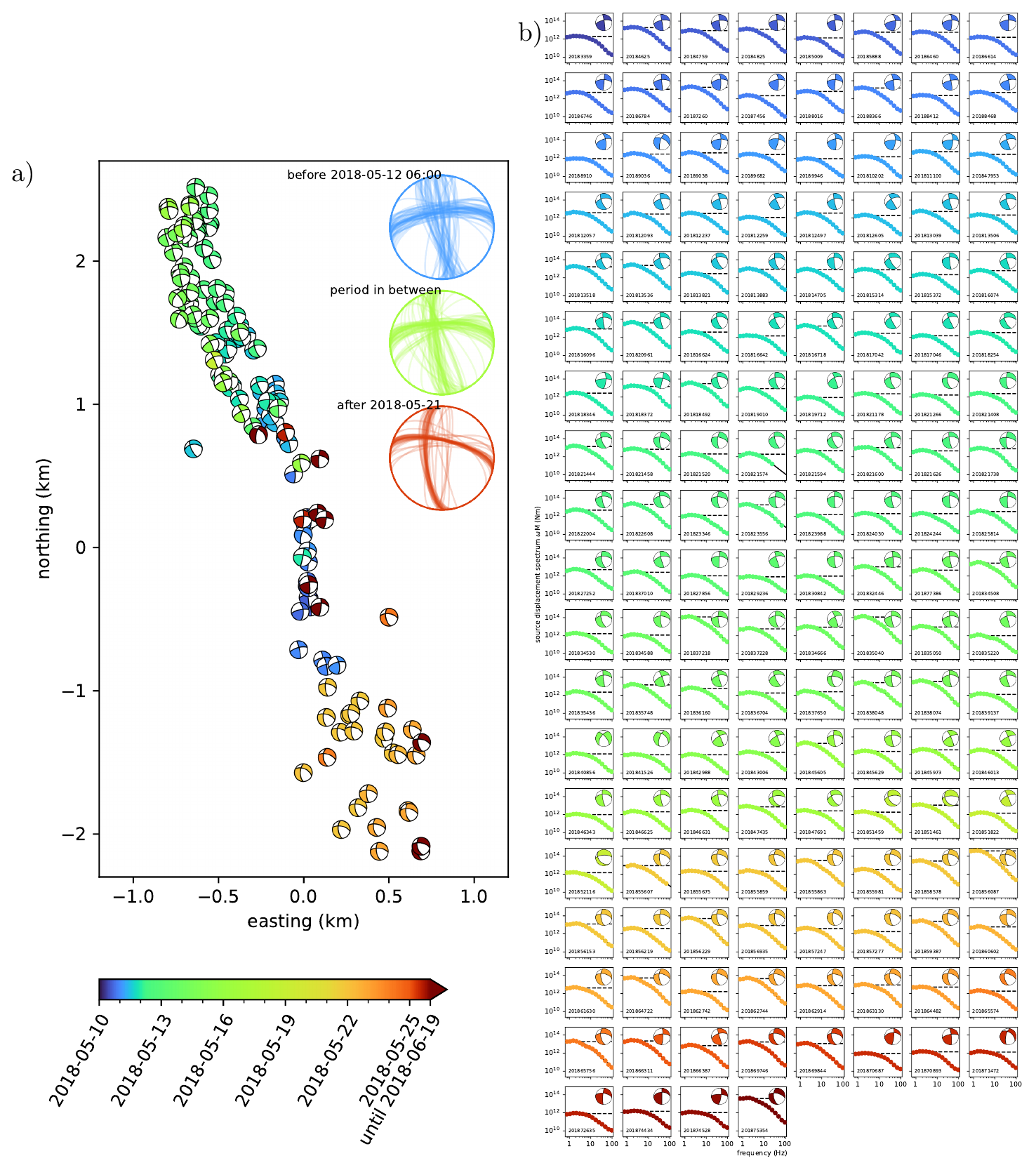}
\caption{a) Focal mechanisms obtained with Grond shown at the catalog locations are color-coded with time.
Coordinates in the map are relative to 50.25°N, 12.45°E. 
The three lower hemispherical projections in the upper right corner show composite nodal lines of earthquakes in three different time periods.\newline
b) Source displacement spectra of the earthquakes calculated with Qopen and color-coded with time. The seismic moments obtained by Qopen are marked with dashed horizontal lines. The radiation patterns of Grond moment tensor solutions are indicated. While both Grond and Qopen invert the data for seismic moment, the two methods target different source parameters -- focal mechanisms (Grond) versus source displacement spectrum (Qopen) -- and are therefore complementary.
}
\label{fig:fm}
\end{figure}

Figure~\ref{fig:fm} displays the two different source parameters obtained from both methods for all analyzed earthquakes. 
The obtained focal mechanisms fall into three families (figure~\ref{fig:fm}a), consistent with the three main fault orientations visible in the double-differences relocated catalog by \cite{Bachura2021} (figure~\ref{fig:eventmap}). Mechanisms for events on the northern and central, deeper faults which occurred mostly before 2018-05-21 show slightly oblique strike slip behavior. The southern, shallower events, active at the end of the sequence are normal faulting with a minor strike-slip component.
The median strike, dip and rake of all fault plane solutions is \SI{171}{\degree}, \SI{79}{\degree} and \SI{-32}{\degree}, respectively (taking into account the bi-modal distribution in strike and rake).
The individual inversions show no significant compensated linear vector dipole (CLVD) component for most events (median error is 0.16). Over all events a small negative CLVD component of -0.045 $\pm$ 0.02 is retrieved (figure~\ref{fig:grond_residuals}), which would be statistically significant if we ignore possible biases.\par

In an earlier study, \cite{Eulenfeld2020} assumed that the earthquakes in the present data set have a similar  focal mechanism as the oblique normal faulting mechanisms of the largest 13 earthquakes with previously known focal mechanism \citep{Plenefisch2019EGU, Bachura2021}. This assumption is in general confirmed by the present analysis, although the waveform inversion can resolve slightly rotated mechanisms on faults with slightly different orientations.
\par
The calculated moment tensors and source spectra are made available online at \url{https://data.pyrocko.org/publications/grond-reports/west-bohemia-2018/} \citep{Eulenfeld2021_grondreport} and \url{https://github.com/trichter/robust_earthquake_spectra} \citep{Eulenfeld2021_sourcecode}.

\subsection{Comparison of magnitudes for envelope and waveform based inversion}
\label{sec:mags}

\begin{figure}
\centering
\includegraphics[width=\textwidth]{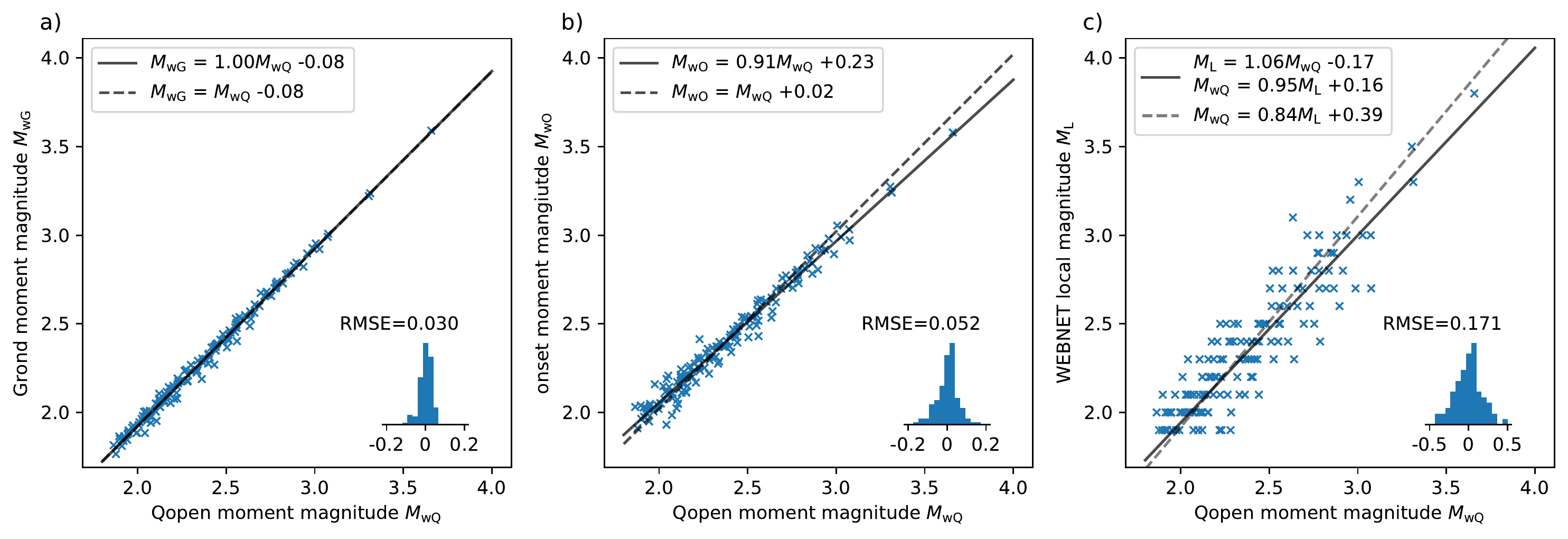}
\caption{Comparison of different magnitude estimates for all processed earthquakes. Moment magnitude from Grond \M wG (panel a), moment magnitude from onsets \M wO (panel b) and WEBNET local magnitude \Ml (panel c) all displayed versus moment magnitude \M wQ estimated with Qopen. The black lines correspond to different linear fits. In panel a and b the continuous line represents a regression between independent variable \M wQ and \M wG, respective \M wO. The slope of the regressions represented by the broken line is fixed at 1 in panels a and b.
In panel c the continuous line represents a regression between independent variable \M wQ and \Ml, the dashed line represents a regression between independent variable \Ml and \M wQ. 

A histogram of the residuals for the regressions corresponding to the continuous lines is displayed in each panel together with the root-mean-square error (RMSE). The regressions in panel a and b show a good linear relationship with a slope around 1 between the different moment magnitude estimates. The RMSE is below 0.1, a typical error for an earthquake magnitude. The lowest RMSE is observed for the regression between the moment magnitudes estimated with Qopen and Grond.
} 
\label{fig:mags}
\end{figure}

In figure~\ref{fig:mags} we compare the moment magnitudes estimated by the envelope inversion (\M wQ from Qopen) with the two waveform based moment magnitude estimates (\M wG from Grond and \M wO from the spectra of onsets) and with WEBNET local magnitude \Ml.
In panel \ref{fig:mags}a the estimates of moment magnitudes from Qopen and Grond are compared for all used events. The regression
$\M wG=1.00\M wQ-0.08$
shows a very good agreement between both estimates with a root mean square error (RMSE) of only $0.030$. Note that the RMSE and the offset of 0.08 are below $0.1$, a typical error obtained when determining an earthquake magnitude. 
The moment magnitudes determined with Qopen and with the spectra of onsets in panel \ref{fig:mags}b show an equally well defined linear relationship with a higher RMSE of $0.052$. The lower estimate of moment magnitudes from onset spectra for the larger earthquakes are noticeable due to their residuals to the regression with fixed slope of~1.
The high fluctuation in the amplitude of low frequency plateaus of the P and S wave spectra at different stations visible in figure~\ref{eq:sds_from_onset}, even after a dedicated correction for radiation pattern and attenuation, could explain obscured seismic moment estimates \M wO. The comparisons in panels \ref{fig:mags}a and \ref{fig:mags}b show that moment magnitudes determined with Qopen are reliable.

In panel \ref{fig:mags}c the WEBNET local magnitude \Ml is compared to the moment magnitude \M wQ determined with Qopen for all used earthquakes.
The regression between assumed independent variable \Ml and \M wQ (dashed line in panel~\ref{fig:mags}c) is given by
$\M wQ = 0.84\Ml+0.39$.
However, we argue that it is better to treat \M wQ as independent variable in the regression, because its residual to moment magnitude is small, while amplitude derived local magnitude generally shows a higher scatter. The regression between assumed independent variable \M wQ and \Ml is given by
$\M wQ = 0.95\Ml +0.16$ (continuous line in panel~\ref{fig:mags}c).
It only uses earthquakes with $\Ml\geq 2.3$ to eliminate any systematic distortion due to the selection of earthquakes with local magnitudes \Ml larger than $1.8$. Both moment magnitude - local magnitude relationships are compared to previous studies determining these relations in figure~\ref{fig:mags_relation} in appendix~\ref{sec:appendix:mags_relation}.

\subsection{Source parameters, scaling between moment magnitude and corner frequency}

\begin{figure}
\centering
\includegraphics[width=0.8\textwidth]{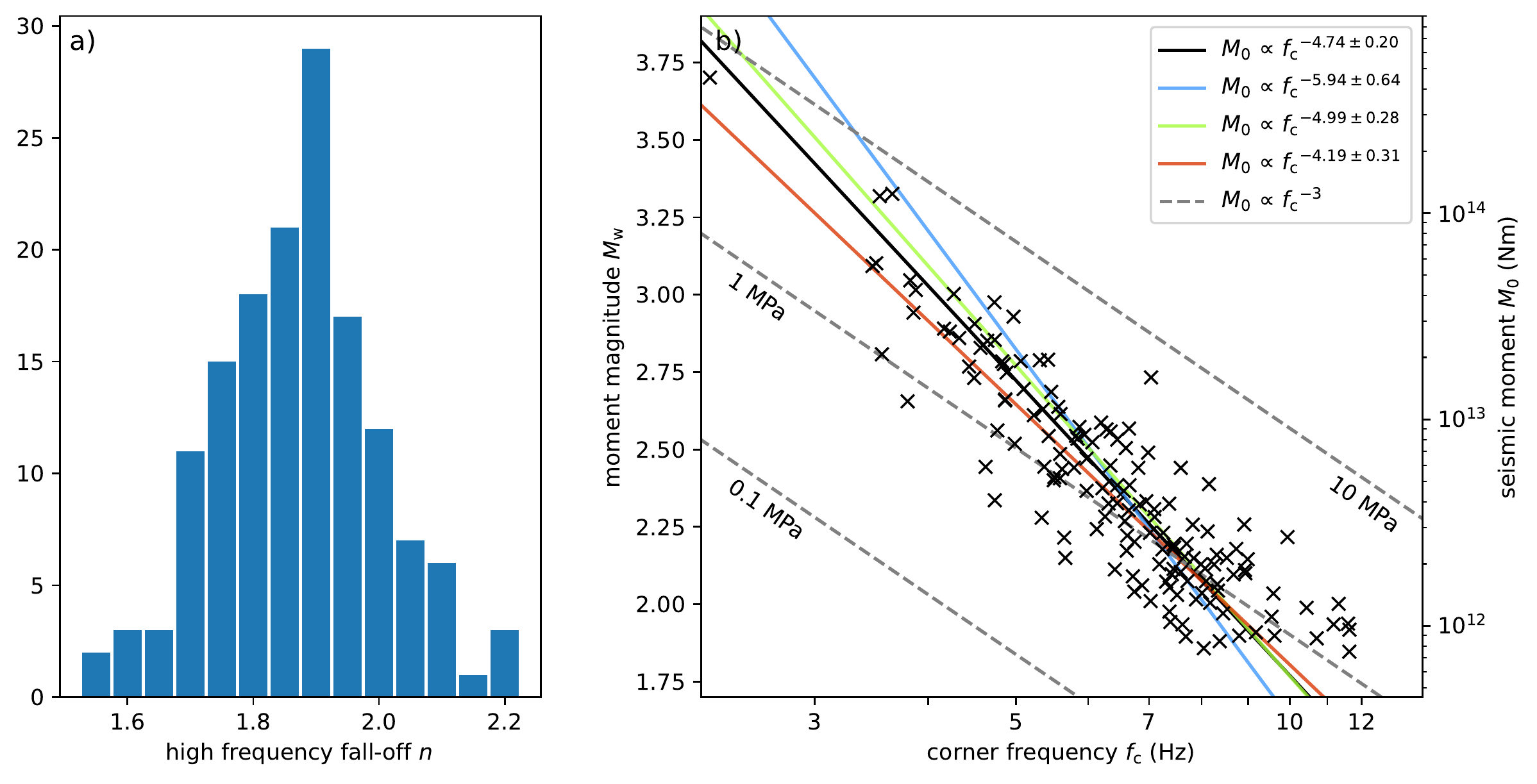}
\caption{a) Histogram of observed high frequency fall-off for all earthquakes.\newline
b) Scaling between moment magnitude and corner frequency, both estimated by the Qopen inversion with fixed high-frequency fall-off.
Constant stress drops of \SI{0.1}{MPa}, \SI 1{MPa} and \SI{10}{MPa} corresponding to a scaling $M_0\propto\fc^{-3}$ are indicated with dashed lines \citep{Madariaga1976}. Linear regressions of subsets of the data (colored lines, for corresponding time periods see figure~\ref{fig:fm}a) and the whole data set (black line) visualize the progression of $M_0-\fc$ relationship with time.
}
\label{fig:fc}
\end{figure}

Qopen allows to quickly determine the earthquake source spectrum and by fitting equation~\ref{eq:sourcemodel}, seismic moment $M_0$, corner frequency \fc, high frequency fall-off $n$ and possibly corner sharpness $\gamma$ can be obtained for each source spectrum. 
We calculated corner frequency and high frequency fall-off for all earthquakes separately while fixing the corner sharpness $\gamma$ at 2. The high frequency fall-offs for all earthquakes show a narrow distribution around a median value of $n{=}1.88$ approaching the classical omega square model (figure~\ref{fig:fc}a). Because of the trade-off between corner frequency and high frequency fall-off we determine the final values for \fc by fixing not only $\gamma$ but also $n$ at its median value. Figure~\ref{fig:fc}b displays the relationship between moment magnitude and corner frequency for the present data set. If we consider the entire duration of the swarm, a linear regression between logarithmic corner frequency and independent variable moment magnitude results in a seismic moment $M_0$ which is approximately proportional to $\fc^{-4.7}$ with a standard error in the exponent of 0.2. When the three time periods given in figure \ref{fig:fm}a are selected for regression, the exponent appears to decrease as time progresses, although this trend can just be resolved because of the increasing errors due to the smaller ensemble size. Again, because of the high variance of data points in figure~\ref{fig:fc}b, the choice of type of regression and independent variable has a high impact on the result. If the logarithmic corner frequency had been chosen as independent variable, $M_0$ would have been on average proportional to $\fc^{-3.8}$.
In figure~\ref{fig:fc}b we additionally display lines of constant stress drop for a circular fault with a rupture velocity of 90\% of the shear wave velocity \citep[equations (24)-(25) for S~waves]{Madariaga1976}. A constant stress drop independent of magnitude corresponds to a proportionality $M_0\propto\fc^{-3}$ for the omega square model \citep{Brune1970}. The present data set does not show a constant stress drop with magnitude in the narrow magnitude range analyzed and negates the self-similarity of studied earthquakes, especially at the beginning of the swarm.

\section{Discussion}

\cite{Michalek2013} calculated source spectra of earthquakes from the 2000 and 2008 West Bohemia swarms from direct body waves (our method in section~\ref{sec:method3}) based on the earlier method of \cite{Hanks1972}. \cite{Michalek2013} use the established procedure to simultaneously invert the body wave spectra for corner frequency \fc and an attenuation $Q$ that is assumed to be constant with frequency \citep{Masuda1982, Hough1999, Edwards2010}. We argue that it is preferable to independently estimate medium properties (attenuation as a function of frequency) and source spectra as in this study and infer source parameters afterwards directly from the obtained source spectra. An additional benefit is that the method does not make any assumption about the underlying source model and we therefore can learn more about earthquakes differing fundamentally from ``common'' earthquakes which are well described by the omega-square model. Deviations from the omega-square model with a higher frequency fall-off $n$ than 2 have been observed earlier, for example in volcanic environment \citep{Ambeh1991} and related to induced seismicity from geothermal stimulation \citep{Eulenfeld2016}.
\cite{Michalek2013} obtained attenuation values $Q^{-1}\ind P$ for P waves of around $\num{4e-3}$ consistent with our attenuation estimate of shear waves at a frequency of \SI 3{Hz}. They report systematically higher corner frequencies than in this study for similar sized earthquakes. While corner frequencies of shear waves used in this study within the envelope are often reported to be higher than corner frequencies of compressional waves used in \cite{Michalek2013} for the same earthquake \citep{Hanks1972}, most of the discrepancy is supposable due to the higher seismic moment calculated in \cite{Michalek2013} for earthquakes with the same local magnitude. The offset in moment magnitude in figure~\ref{fig:mags_relation} is around 0.6 at local magnitude 2 corresponding to a factor of 1.6 in corner frequencies for $M_0\propto\fc^{-4.7}$.
This discrepancy propagates to estimates of stress drop: \cite{Michalek2013} report a stress drop around \SI{10}{MPa} for earthquakes with a seismic moment of \SI{e13}{Nm} compared to \SI{1.8}{MPa} in this study (figure~\ref{fig:fc}b).
The large negative exponent of $-4.7$ in the $M_0-\fc$ relationship for the 2018 swarm is unexpected. \cite{Michalek2013} report  a similar large exponent also for the swarms in 2000 and 2008, although their database was much smaller.
Because the three different stages of the 2018 swarm do not show an overlap in space and time (figure~\ref{fig:eventmap}), the stress drop variation correlate in time as well as with fault segments.
A possible reason for variations in stress drop may be the tectonic setting or the orientation of the faults. For instance, \cite{Goebel2015} studied the tectonic and structural control of stress drop heterogeneity in the San Gorgonio area of the San Andreas fault system, southern California. While normal faulting earthquake were generally associated with lower stress drops at about \SI 4{MPa},  thrust faulting earthquakes showed higher stress drops at about \SI 6{MPa}. Variations of stress drop between about 1 and \SI{10}{MPa} were found on neighboring faults with different tectonic locking. \cite{Goebel2015} concluded that stress drop is approximately inversely correlated to slip rates on the fault systems which might explain our results although the variability in source mechanism is relatively small.
\par 

Another control of stress drop variability is pore pressure. \cite{GoertzAllmann2011} estimated stress drop variability of injection-induced earthquakes at the Basel geothermal site at about 4 km depth of two orders of magnitudes. They could correlate the stress drop with spatial pore pressure changes. However, a systematic increase of stress drop with magnitude was not evident.  
\cite{Lengline2014} found stress drop changes of a factor of 300 for fluid-induced earthquakes occurring at the same locality at different times, and associated this with the temporal variation of pore pressure. Studies of previous earthquake swarms in NW Bohemia had yielded evidence for local fluid motion and temporal pressure variations in the focal area. Using a relative method, \cite{Dahm2013} found strong short-term decreases of the $v\ind P/v\ind S$ ratios at the beginning of swarms and attributed this to the intrusion of gases (e.g.\ CO2). One of the hypotheses developed to explain the swarm earthquakes in NW Bohemia are magmatic intrusions associated with the release of gas bubbles, which together can explain the size and shape of the seismicity and the temporal patterns and migrations \citep{Dahm2008}.
The gradual change in the stress drop for larger earthquakes of the same magnitude (figure~\ref{fig:fc}b) may be explained by pore pressure, which is enhanced during the beginning phase of the swarm and declining to more normal values with progressing time. Earthquakes triggered under high pore pressure may involve shearing and opening and thus a reduced contact area of fault planes. As rupture speed is slowed down by friction, the source duration is expected to be shorter in the beginning phase of the swarm, when pore pressure in enhanced, compared to the ending phase when pore pressure is recovering to background.
The transition of the 2018 activity from swarm stage to mainshock/aftershock stage (figure\ref{fig:eventmap}b) might be related.


\par

As mentioned earlier, our approach allows to estimate seismic moment and moment magnitude from the independently determined source spectra. Seismic moments can also be estimated with a related approach using coda waves with a subsequent calibration to larger earthquakes with known moment tensor from the inversion of direct arrivals \citep[e.g.][]{Rautian1978, Mayeda1996, Mayeda2003, Holt2021}. Our method does not need any calibration since it is based on a physical model and can therefore be used in regions with a lack of large earthquakes. Additionally, our method can be used to estimate earthquake spectra and seismic moments in high scattering environments (e.g. volcanoes), for which traditional methods of moment tensor inversion based on direct waves fail, because of to the lack of impulsive onsets. 

A key benefit of our method is that it does not rely on prior knowledge on intrinsic and scattering attenuation values, and relative site amplification factors, especially since these are often unknown for newly monitored regions. We determine and fix them during our inversion scheme along with the source parameters. Their values and frequency dependence can lead to additional geological insights and they serve as valuable inputs for follow-up methods like moment tensor inversion. Site amplification for selected WEBNET stations is discussed in appendix~\ref{sec:appendix:sites}.
Prior knowledge of site amplification at specific stations can be taken into account and enhance the determined source spectra, especially at high frequencies.


\section{Conclusions}

We presented a new method to calculate earthquake source spectra with the Qopen method of full envelope inversion. Earthquake source parameters -- seismic moment, corner frequency, high frequency fall-off and moment magnitude -- can be robustly determined in an automatic manner with the presented inversion scheme. 
Source spectra and moment magnitudes were estimated without tuning the estimates to previously characterized earthquakes. The results agree well to moment magnitudes estimated with dedicated Grond package for moment tensor inversion. Both methods, full envelope inversion and moment tensor inversion, should be regarded as a complement as both target different source parameters.

Our work corroborates previous results that established scaling relationships between the seismic moment and the size of the rupture surface are not fulfilled in swarm earthquakes in NW Bohemia. Temporally and spatially variable pore pressures and possible magmatic intrusions are addressed as the cause of the breakdown of scaling.

\appendix

\section{Relationship between local magnitude and moment magnitude for West Bohemia}
\renewcommand\thefigure{\thesection\arabic{figure}}
\setcounter{figure}{0}
\label{sec:appendix:mags_relation}
\begin{SCfigure}
\centering
\includegraphics[width=0.65\textwidth]{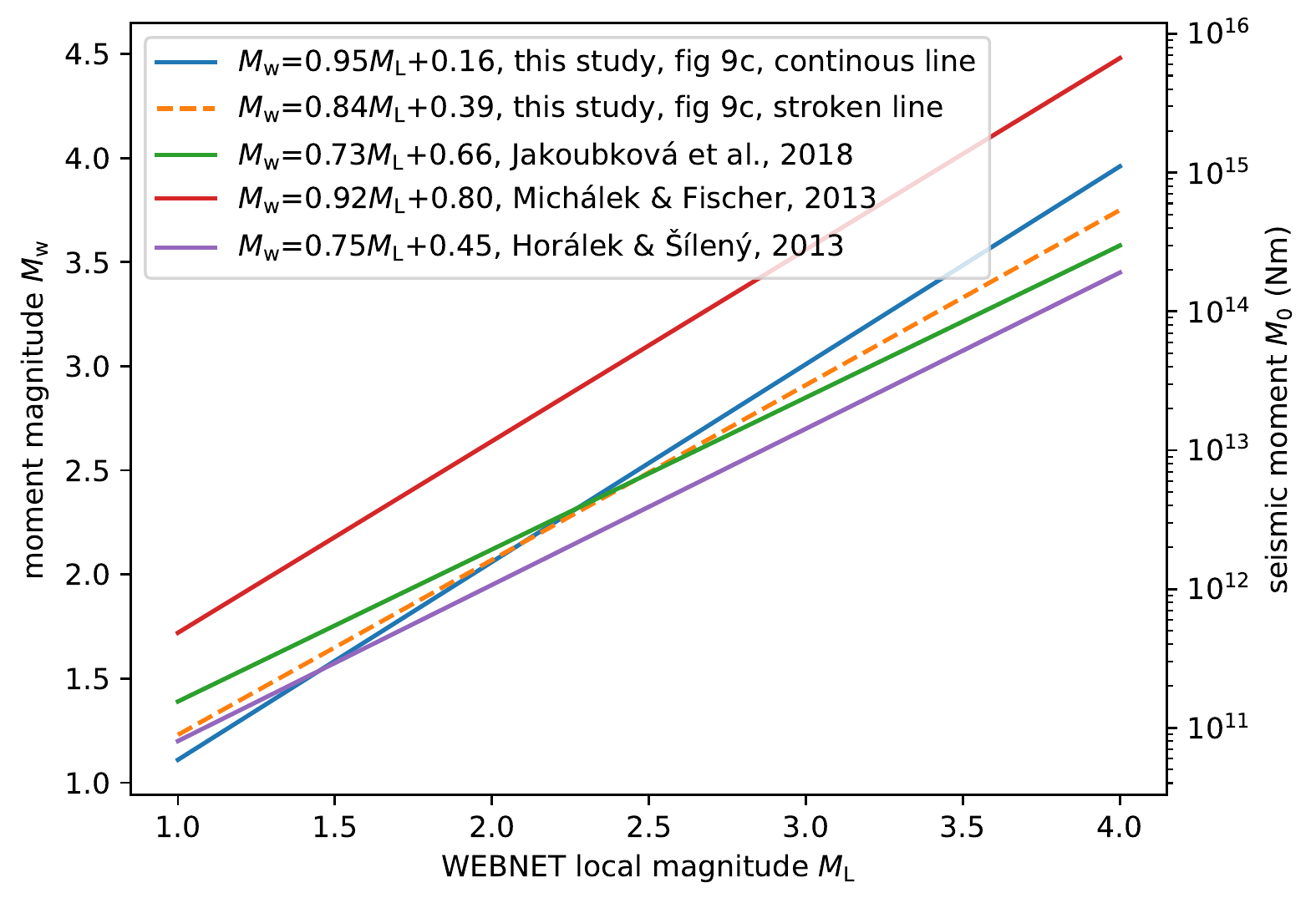}
\caption{Different relationships between moment magnitude \Mw and WEBNET local magnitude \Ml.\vspace{2cm}}
\label{fig:mags_relation}
\end{SCfigure}

\nocite{Horalek2013}

In figure~\ref{fig:mags_relation} we compare different relationships between moment magnitude and WEBNET local magnitude with our fits in figure~\ref{fig:mags}c. The preferred estimate of this study treating the moment magnitude as independent variable $\Mw{=}0.95\Ml{+}0.16$ (blue line) has a similar trend as the most recent estimate from \citet[green line]{Jakoubkova2018}. The slope of our estimate is slightly steeper and is comparably to the slope of \citet[red line]{Michalek2013} with an offset of 0.6 at a local magnitude of 2. When performing the fit between the two magnitude scales in a more conservative way by treating the local magnitude as an independent variable (orange dashed line) the agreement to the latest study from \cite{Jakoubkova2018} is even stronger.

\section{Energy site amplification for selected WEBNET stations}
\setcounter{figure}{0}
\label{sec:appendix:sites}
\begin{figure}
\centering
\includegraphics[width=\textwidth]{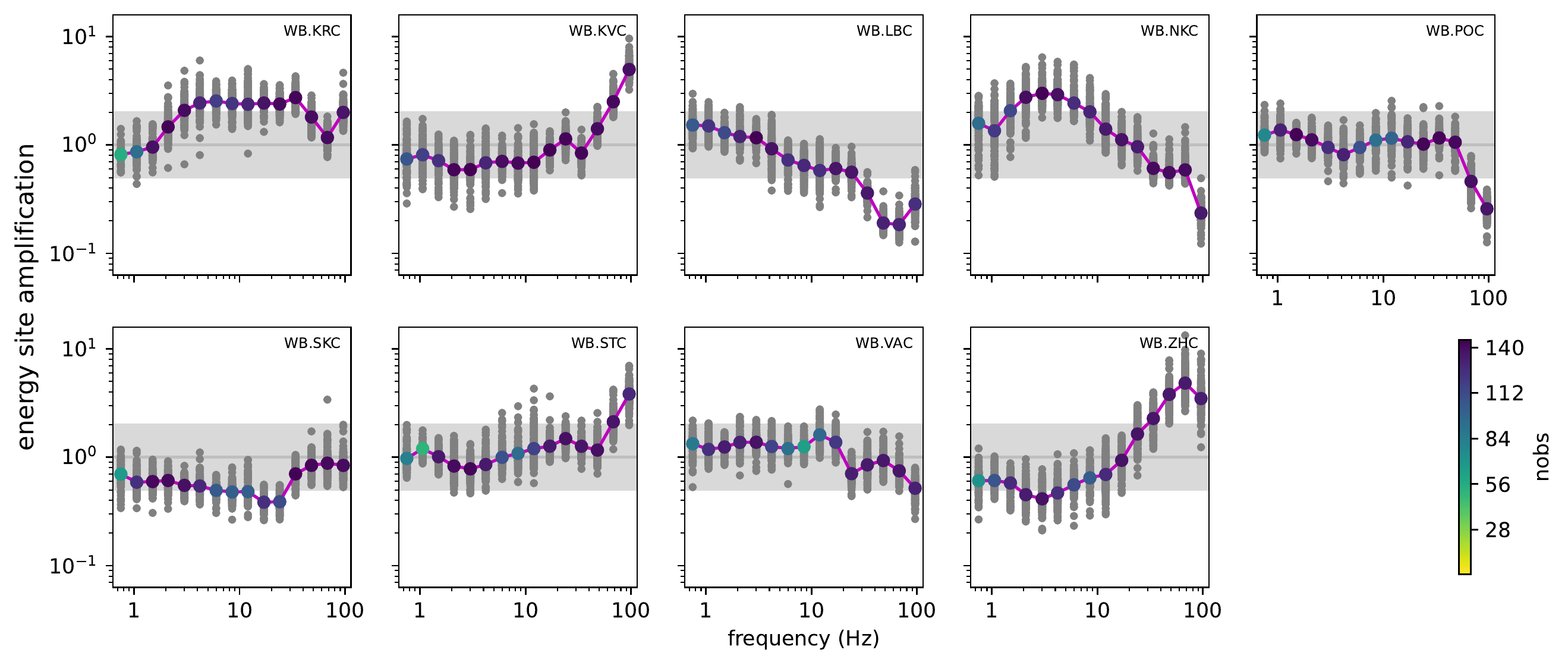}
\caption{Energy site amplification as a function of frequency for the used WEBNET stations. Grey dots depict site amplification measured for a single earthquake. The geometric averages are marked as circles, whereat the color represents the number of observations (i.e.\ number of earthquakes). Note that the geometric mean of all observations including all stations is itself fixed at 1 in each frequency band. The average at 1 is marked with horizontal gray lines. The light gray boxes span amplifications between 0.5 and 2.}
\label{fig:sites}
\end{figure}

Site amplification factors as a function of frequency at the different used WEBNET need to be determined within the inversion of the envelopes for earthquake source spectra described in section~\ref{sec:method1}.
Because the site amplification is mainly influenced by the coda envelope due to its longer duration compared to the envelope of the direct S~wave, influence of source radiation pattern on the site amplification is expected to be of no or minor significance. Figure~\ref{fig:sites} displays the obtained energy site amplification factors (corresponding to the square of amplitude site amplification factors) for the 9 used WEBNET stations. The geometric mean of the site amplification factors of the different stations are fixed at 1 for each frequency band. In general the site response show only small differences between the different stations illustrating the high quality of the selected sites. The highest variation up to a factor of 5 from the average is visible only for high frequencies above \SI{50}{Hz}. Stations KVC, STC and ZHC show a high site amplification at high frequencies. Accordingly site amplification of stations LBC, NKC and POC is lower than average for high frequencies.
Other noteworthy deviations from average amplifications embrace KRC, where amplifications exceed 2 in the frequency band \SI 3{Hz} to \SI{30}{Hz}; NKC, where amplifications exceed 2 in the frequency band \SI 2{Hz} to \SI 8{Hz}; SKC, where amplifications are less than 0.5 in the frequency band \SI{15}{Hz} to \SI{30}{Hz} and ZHC, where amplifications are less than 0.5 in the frequency band \SI 2{Hz} to \SI 4{Hz}.

\paragraph{Data and resources}
\begin{footnotesize}
All data used in this study can be downloaded at \url{https://doi.org/10.5281/zenodo.3741464} \citep{DatasetEQSwarm2018}.
We used Qopen version 4.1 available at \url{https://github.com/trichter/qopen} \citep{Qopen}.
Additional data processing and plotting was performed with the libraries Grond, ObsPy, Pyrocko, NumPy and matplotlib \citep{Grond, Megies2011, Pyrocko, SciPy, Hunter2007}. 
This article can be reproduced with the source code provided at \url{https://github.com/trichter/robust_earthquake_spectra} \citep{Eulenfeld2021_sourcecode}. Results of the Qopen method and the spectrum method can be downloaded from the mentioned link. Additionally, results from the moment tensor inversion can be browsed and downloaded at \url{https://data.pyrocko.org/publications/grond-reports/west-bohemia-2018/} \citep{Eulenfeld2021_grondreport}.
\end{footnotesize}

\paragraph{Acknowledgments}
\begin{footnotesize}
We thank Martin Bachura for providing data from WEBNET seismic stations \citepalias{NetworkWebnet}
and for providing the double difference earthquake catalog \citep{Bachura2021}. Comments from two anonymous reviewers and associate editor Adrien Oth helped to improve the manuscript.
\end{footnotesize}

\catcode`\^^M=5

\begin{footnotesize}
\setlength{\bibsep}{1.5ex}

\end{footnotesize}

\end{document}